# Selected results from the static characterization of edgeless *n*-on-*p* planar pixel sensors for ATLAS upgrades

Gabriele Giacomini[a*], Alvise Bagolini[a], Marco Bomben[b], Maurizio Boscardin[a], Luciano Bosisio[c], Giovanni Calderini[b,d], Jacques Chauveau[b], Alessandro La Rosa[e], Giovanni Marchiori[b] and Nicola Zorzi[a]

[a] *Centro per i Materiali e I Microsistemi, Fondazione Bruno Kessler (CMM-FBK), via Sommarive 18, I-38123 Trento, Italy*
 *E*-mail: *ggiacomini@fbk.eu*

[b] *Laboratoire de Physique Nucleaire et de Hautes Énergies (LPNHE), 75005 Paris, France*

[c] *Università di Trieste, Dipartimento di Fisica and INFN, I-34127 Trieste, Italy*

[d] *Dipartimento di Fisica E. Fermi, Università di Pisa, and INFN Sez. di Pisa, Pisa*

[e] *Section de Physique (DPNC), Université de Genève, Genève, Switzerland*

ABSTRACT: In view of the LHC upgrade for the High Luminosity Phase (HL-LHC), the ATLAS experiment is planning to replace the Inner Detector with an all-Silicon system. The *n*-on-*p* bulk technology represents a valid solution for the modules of most of the layers, given the significant radiation hardness of this option and the reduced cost. There is also the demand to reduce the inactive areas to a minimum. The ATLAS LPNHE Paris group and FBK Trento started a collaboration for the development on a novel *n*-on-*p* edgeless planar pixel design, based on the deep-trench process which can cope with all these demands. This paper reports selected results from the electrical characterization, both before and after irradiation, of test structures from the first production batch.

---

[*] Corresponding author.

# Contents

# 1. Introduction

In the next decade the CERN Large Hadron Collider (LHC) should be upgraded to the so-called high luminosity LHC (HL-LHC) [1], capable of a luminosity of 5 $10^{34}$ cm$^{-2}$ s$^{-1}$, to extend its physics reach. By then the ATLAS experiment will be equipped with a completely new pixel detector. The innermost layer of the new pixel detector will integrate a fluence of about $10^{16}$ (1-MeV) $n_{eq}/cm^2$ for an integrated luminosity of 3000 fb$^{-1}$. These harsh conditions demand radiation-hard devices and a finely segmented detector to cope with the expected high occupancy. Furthermore, the new pixel sensors will need a high geometrical acceptance, and the geometric inefficiency are required to be less than 2.5% [2]. In conventional sensor designs many Guard Rings (GRs), essential to improve the voltage-handling capability of the sensor, surround the active region, so that a relatively large dead area at the edge of the sensor results. One way to reduce or even eliminate this insensitive region along the device periphery is offered by the "active edge" technique [3], in which a deep vertical trench is etched along the device periphery throughout the entire wafer thickness, thus performing a damage free cut. The trench is then heavily doped, extending the ohmic contact on the backside to the lateral sides of the device: the depletion region can then extend to the edge without causing a large current increase.

    This is the technology we have chosen for realizing *n*-on-*p* pixel sensors with reduced inactive zone. Details on the technology can be found in [4]. FBK has already experience in the fabrication of active-edge sensors [5]. With respect to *p*-on-*n* productions, a complication in the process arises from the fact that *n*-electrodes would be shorted by the electron accumulation layer present at the Si/SiO$_2$ interface unless a *p*-type compensating implants is present. In the first production run at MtLab of FBK, we tested different solutions for achieving such isolation: for different process splittings, two different doses of *p*-spray (i.e. an unpatterned, uniform, low-dose *p*-implant) of $3X10^{12}$ cm$^{-2}$ and $5X10^{12}$ cm$^{-2}$ and presence or absence of *p*-stops (i.e. patterned, high dose *p*-implants in between the pixels) were used. Studies performed with TCAD simulation tools helped in defining the layout and in making a first estimation of the charge collection efficiency expected before and after irradiation for different geometries [6].

    In Section 2 some selected results about the electrical characterization of some pixel test structures will be presented. In Section 3 the automatic measurements performed on the FE-I4 compatible pixel sensors will be shown. Conclusion are drawn in Section 4.

## 2. Electrical characterization of pixel test structures.

The largest part of a wafer is dedicated to 9 FE-I4 compatible pixel sensors, which consist of an array of 80X336 pixels, at a pitch of 250 μm X 50 μm. The nine sensors differ for the GR structures [4]. On the wafer periphery several test structures were placed, which can anticipate some of the main sensor performances. One of the parameters to be extracted from the testing was the breakdown voltage, which is expected to vary for the different designs, namely, number of GRs and pixel-to-trench distance. The test-structures for the measurement of this parameter, which are described in [7], consist in an array of FE-I4-like 6X30 pixels, shorted by narrow metal links, in order to be contactable by a single probe. They differ for the number of GRs as well as for the pixel-to-trench distance (Fig. 1).

Another important parameter is the pixel capacitance. To measure it, we inserted specific test patterns, consisting of a contactable FE-I4 pixel surrounded by concentric rings of contactable FE-I4 pixels [7]. For test purposes, they come in two variants: with and without metal field plate (FP) over the *n*-plus implant (although all pixels have FP in the FE-I4 detectors).

Some of these test structures have been irradiated at the Triga nuclear reactor of JSI of Ljubljana [8] (at a fluence of 2.5 X $10^{15}$ (1-MeV) $n_{eq}/cm^2$, which is the expected dose the pixel will integrate in ATLAS). An example of the IV after irradiation will be given in Section 2.1.1.

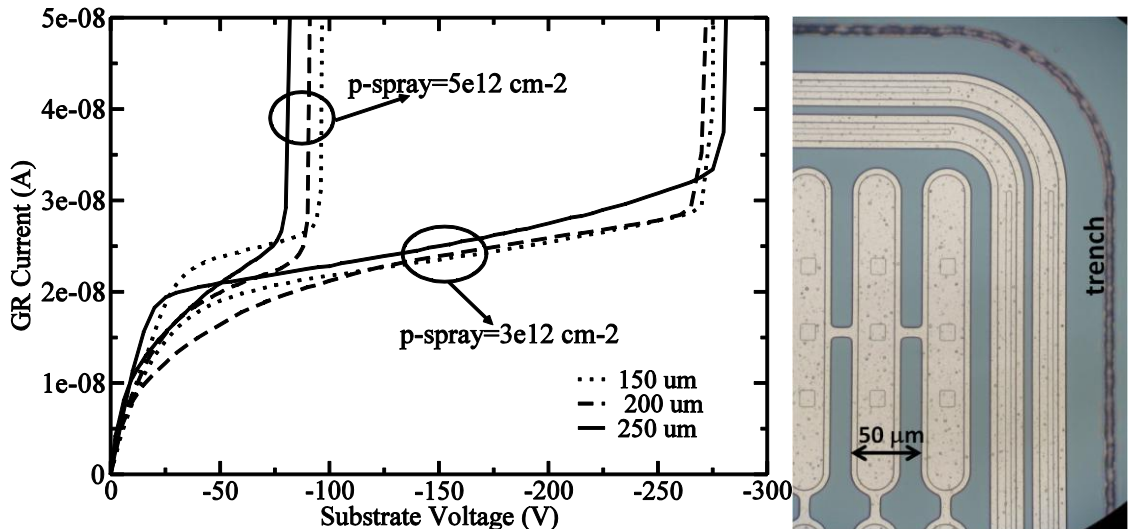

*Fig. 1: I-Vs of the inner Guard Ring of some un-irradiated pixel test structures belonging to two wafers, one with a p-spray dose of $3X10^{12}$ $cm^{-2}$ and the other of $5X10^{12}$ $cm^{-2}$. The pixel test structures consist of 6X30 FE-I4- like pixels surrounded by 3 GRs (the innermost of them is grounded) and the pixel-to-trench distance is as reported in the legend. The microscope picture shows a corner of such a test structures, but featuring two GRs.*

### 2.1 Breakdown voltage measurement

In Fig. 1 there are reported the current-voltage characteristics for the test structures for the measure of the $V_{BD}$, featuring three GRs (the innermost is grounded, and its current plotted, while the external two are floating) and different pixel-to-trench distances (150, 200 and 250 µm). In Fig. 1 there are superimposed the curves of the same test structures coming from a wafer having a p-spray dose of $3\times10^{12}$ cm$^{-2}$, and another wafer having a *p*-spray dose of $5\times10^{12}$ cm$^{-2}$. The pad current, not plotted, shows no breakdowns and it is independent of the pixel-to-trench distance, while its value is compatible with the one measured on test diodes. As can be seen, the breakdown voltage has a strong dependence on the *p*-spray dose, while it is quite independent of the pixel-to-trench distance. The steep rise of the GR current allows us to determine the $V_{BD}$ for the different number of GRs, as summarized in Table 1.

Table 1: average $V_{BD}$, as measured from test structures with different pixel-to-trench distances, for different GR structures and *p*-spray doses.

|        | $V_{BD}$ (V) <br> *p*-spray dose ($3\times10^{12}$ cm$^{-2}$) | $V_{BD}$ (V) <br> *p*-spray dose ($5\times10^{12}$ cm$^{-2}$) |
|--------|--------|--------|
| No GR  | 90  | 40 |
| 1 GR   | 120 | 50 |
| 2 GRs  | 200 | 75 |
| 3 GRs  | 250 | 90 |
| 5 GRs  | 300 | 90 |
| 10 GRs | 300 | 95 |

### 2.1.1 $V_{BD}$ on irradiated devices

The same electrical characterization has been performed on irradiated devices. After the irradiation, the samples have experienced a limited annealing at room temperature (during the shipment), then they have been stored at -20 °C, while the measurements have been performed at 0 °C. An example of IVs of irradiated structures can be seen in Fig. 2: as expected, the irradiation increased the $V_{BD}$ and, despite the very limited number of GRs (only 3 in this example) and the proximity of the trench, a $V_{BD}$ in excess of 500 V has been achieved. More detailed measurements will follow in a subsequent publication.

### 2.2 Pixel capacitance

Capacitance measurements, for pixels with and without FP, for all the isolation methods (different *p*-spray doses and presence/absence of *p*-stop), as a function of the frequency (for several bias voltages) and as a function of bias voltages (for several frequencies) have been performed.

In Fig. 3, the interpixel capacitance vs voltage of the central pixel toward the first neighbors (first ring of 8 pixels) is plotted. Such measurement has been found almost independent of the frequency: in the plot, only the data for 100kHz are reported. The capacitance relative to pixels without FP (dotted lines) is smaller than the capacitance of the corresponding structure with FP, while structures with the higher *p*-spray dose exhibit a larger capacitance than those with a smaller dose. These differences are all expected but they are of small concerns (in the 10fF range). A different picture arises when considering the interpixel capacitance between the central pixel and all the other pixels (first and outer rings), as shown in Fig. 4. For a *p*-spray dose of $3\times10^{12}$ cm$^{-2}$, the capacitance is practically independent of the bias voltage as well as of the frequency (here again, there is a little difference between structures featuring the FP or not). This suggests that the interface below the FP is depleted from the *p*-spray, and thus there is no long-range coupling between the pixels. The matter is different for a *p*-spray dose of $5\times10^{12}$ cm$^{-2}$. Here the capacitance is by far larger and there is a strong voltage dependence. Also, the capacitance of the structure without FP is much lower. These facts suggest a strong and long-range coupling between the pixels, aided by the resistive *p*-spray. By increasing the bias voltage, the *p*-spray gets more and more depleted under the FP, decreasing the capacitance.

Even a pixel capacitance of 150 fF is not so high as to prevent the achievement of good results in terms of input capacitance. For instance, 3D pixels (read-out by the same FE-I4 chip) have a capacitance of about 200 fF and nonetheless can achieve a noise level of only 150 e. r.m.s [9].

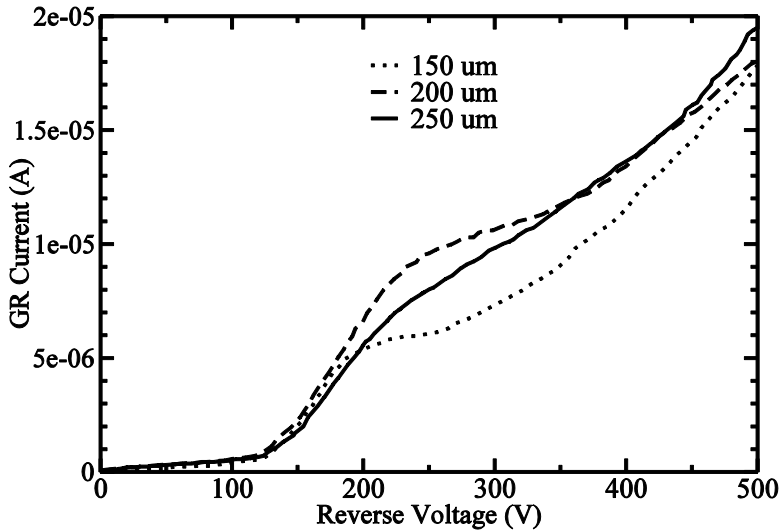

*Fig. 2: I-V of the Guard Ring of some irradiated pixel test structures belonging to a wafer with a p-spray dose of $3\times10^{12}$ cm$^{-2}$, having 3GRs and the pixel-to-trench distance as reported in the legend.*

The single pixel capacitance towards the backside can be estimated below 1 fF/pixel from the parallel plane approximation, therefore cannot be directly measured with enough accuracy. To evaluate this parameter, we measured the total capacitance of all the 117 pixels of the test structure toward the backside. Fig. 5 shows the obtained results for three different isolation conditions as a function of the frequency. The capacitance at low frequency accounts for the coupling with the interfacial resistive layers. The RC associated with this network of distributed resistances and capacitances is related to the inverse of the frequency at which a bump in the dissipation factor occurs and is strongly dependent on the presence or not of the p-stop. The high-frequency asymptotic value of the capacitance (i.e. the value at the operational point for $f >$ 1MHz) is the direct capacitance of the pixels toward the bulk, which does not depend on the interfacial resistive layers: it is about 50 fF for the 117 pixels, i.e. about 0.5fF/pixels, compatible with the parallel plane approximation.

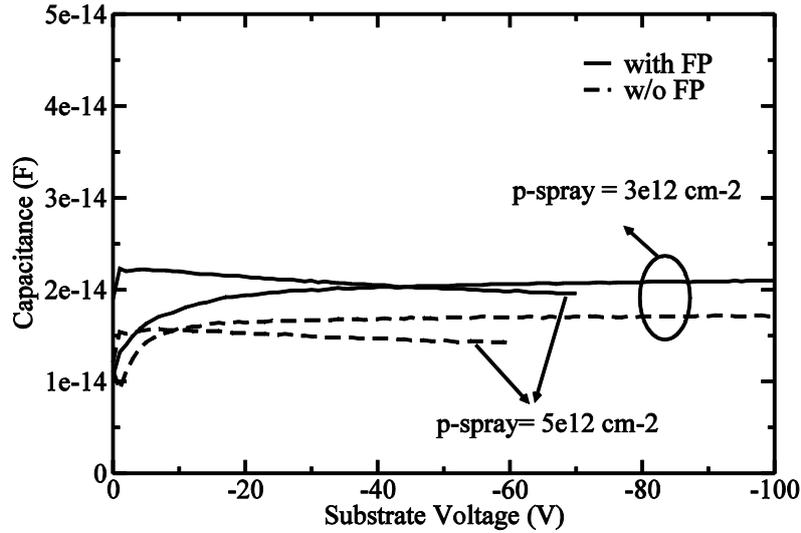

*Fig. 3: Interpixel capacitance between the central pixel and the first neighbours, for structures coming from wafers having different p-spray doses. As in the legend, solid lines refer to pixel featuring the FP, while dotted lines refer to pixel without FP. Frequency = 100 kHz. The pixel test structures consist of 9X13 FE-I4-like pixels, organized in rings.*

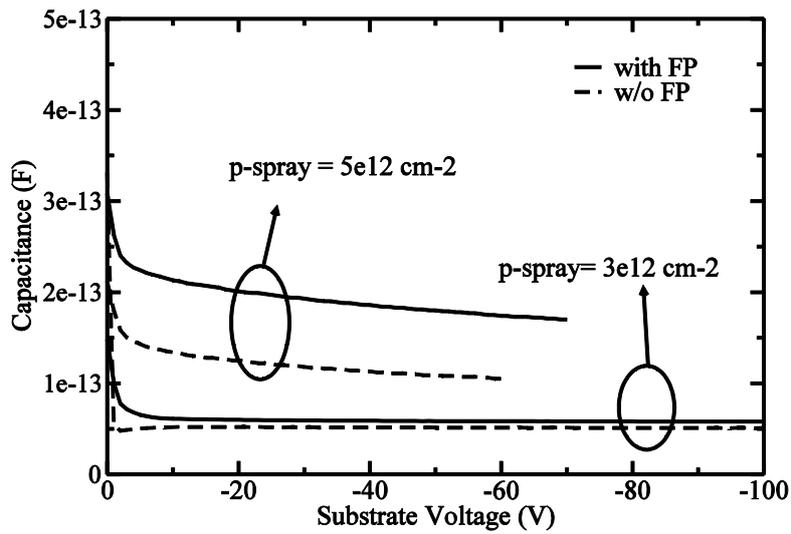

*Fig. 4: Interpixel capacitance between the central pixel and all the neighbours, for structures coming from wafers having different p-spray doses. Solid and dotted lines as in Fig. 3. Frequency = 100 kHz.*

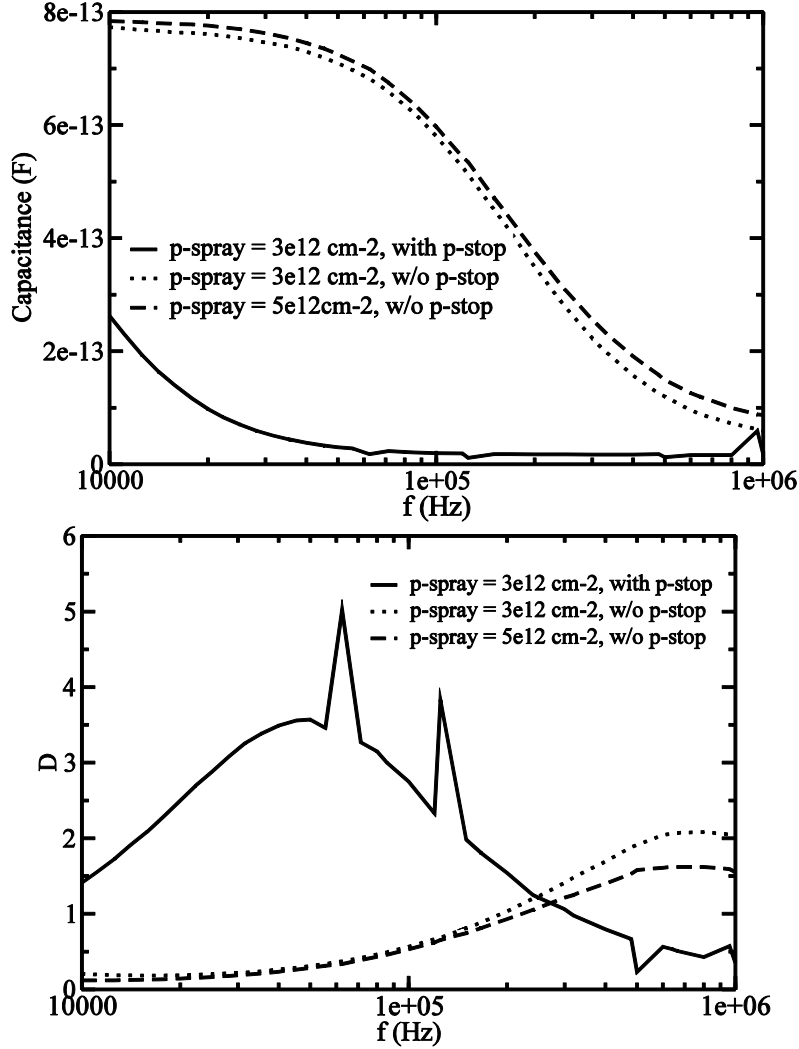

*Fig. 5: (top) Capacitance of all the pixels toward the back-side, for different p-spray/p-stop implants, as specified in the legend.(bottom) Dissipation factor relative to the same measurement. $V_{bias}=30V$.*

## 3. IV on FE-I4 sensors

To select functional FE-I4 compatible pixel-sensors to be bump-bonded to the read-out chips, an electrical test at the wafer level is highly desirable. If the testing procedure is straightforward as far as strip sensors are concerned, this is not true for pixel sensors, because it is not practical to connect so many pixels with a probe card. Thus, the same approach used for 3D pixel sensors [10] has been used also here: i.e. a temporary metal was deposited above the passivation layer. The temporary metal contacts the metal of the underneath pixel through the opening of the passivation layer (Fig. 6), where the under-bump metal first and then the indium bump-ball will be placed. In principle all the pixels can be connected together so as to measure the sensor total current with a single electrode. However, we decided to have a segmented temporary metal in order to get some insight about the spatial distribution of possible defects. Therefore, a sensor needs 80 metal stripes, each shorting a row of 336 pixels. During the IV measurement, only about 10 stripes are grounded by the probecard, and their current read-out, while the other rows are unbiased and thus stay at the bulk voltage, provided by an additional probe contacting the bias tab. After the measurement, the metal is removed by a simple wet etch, which cannot spoil the electrical characteristics just measured. Similarly to the 3D sensors, in which the ohmic columns touch the *p*-spray on the *n*-side, here the *p*-spray is at the same voltage of the back, so as to make the full applied voltage to drop at the pixel border. The break-down voltage we expect is then the one measured on the pixel test structures having no GRs (about 90 V and about 40 V for a *p*-spray dose of $3X10^{12}$ cm$^{-2}$ and $5X10^{12}$ cm$^{-2}$, respectively). In Fig. 7 there is an example of such measurements. The sum of the 80 currents (one for each stripe) of the 9 pixel sensors coming from one wafer (having *p*-spray dose of $3X10^{12}$ cm$^{-2}$) is plotted. Three sensors are defective since their break-down voltage is well-below the expected 90 V. All the other sensors are fine and are good candidate for the bump-bonding. Notice the large current difference between the only sensor without GRs and the others (featuring GRs). This is due to the fact that, if present, GRs are biased by punch-through by the nearby pixels, creating a quite large depletion region, whose generated leakage current is collected by the pixels. This amount of current is summed 80 times in the plot. In Fig. 8, the 80 IVs of one of a defective sensor of Fig.7 are shown: as can be seen, only one row is defective, while the others are absolutely fine. We can deduce that only few pixels are defective in this row but this is enough to spoil the electrical properties of the whole sensors. The same thing has been observed with 3D sensors.

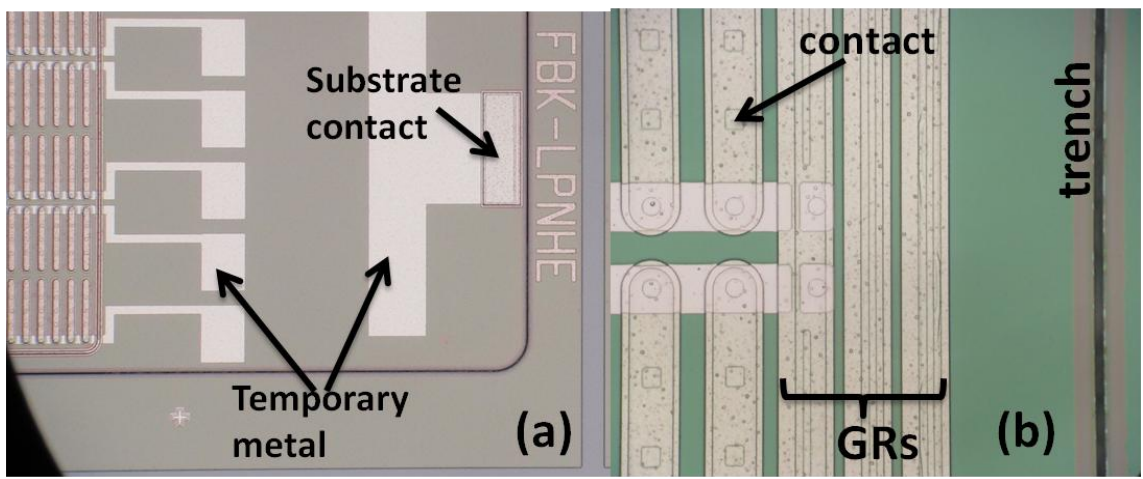

*Fig. 6: (a)Temporary metal on an FE-I4 compatible pixel sensors, showing the stripes shorting the pixels and the bus for contacting the substrate. (b) Larger Zoom of the temporary metal on few pixels.*

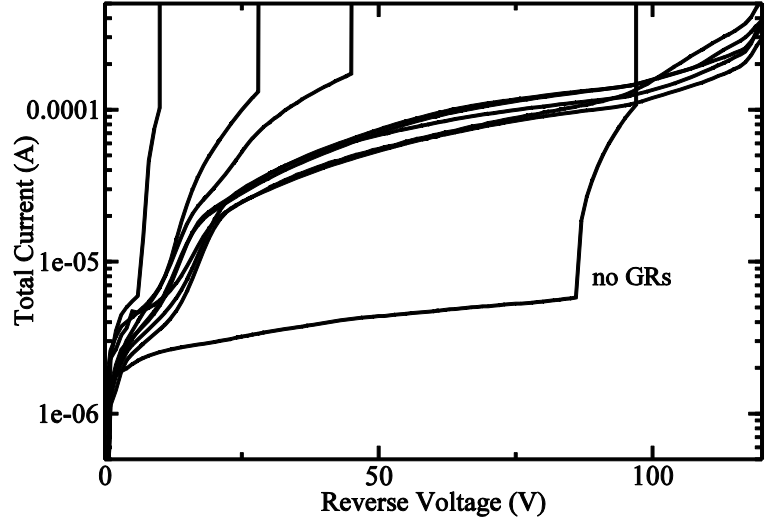

*Fig. 7: Sum of the currents of the 80 stripes, for each of the 9 FE-I4 sensors of a wafer. P-spray = $3 \times 10^{12}$ cm$^{-2}$.*

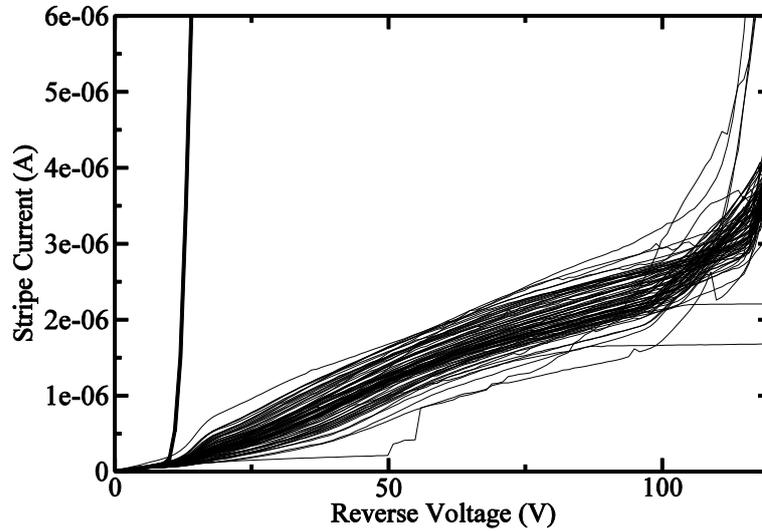

*Fig. 8: Currents of the 80 stripes of a sensor which was found defective with the total IV of Fig. 7. Only one stripe out of 80 is defective.*

## 4. Conclusions and outlook

In view of the upgrade of the ATLAS Inner Detector for HL-LHC runs, FBK Trento and LPNHE Paris are developing new planar *n*-on-*p* pixel sensors, characterized by a reduced inactive region, obtained by means of a trench: the "active edge" approach. The measurements performed on test structures, including capacitance- and current-voltage characteristics, show that it is possible to operate the full sensors successfully, well in over-depletion. Functional tests of the FE-I4 pixel sensors with radioactive sources, before and after irradiation, and eventually in a beam test, after having bump bonded a number of pixel sensors to the FE-I4 read-out chips, will follow.

**Acknowledgments**

We acknowledge the support from the MEMS2 joint project of the Istituto Nazionale di Fisica Nucleare (INFN) and Fondazione Bruno Kessler.